\documentclass[5p,review,number]{elsarticle}

\usepackage{hyperref}
\usepackage[T1]{fontenc}
\usepackage[utf8]{inputenc}
\usepackage{textcomp} 
\usepackage{lmodern} 
\usepackage{amsmath}
\usepackage{amsfonts}
\usepackage{breakurl} 
\usepackage[switch]{lineno} 
\graphicspath{{./figures/}}
\biboptions{sort&compress}
\usepackage{subcaption} 
\usepackage{booktabs}
\graphicspath{{./figures/}}
\usepackage{dcolumn}
\usepackage{multirow}
\usepackage{adjustbox}
\usepackage{bm}
\usepackage{textgreek}

\hypersetup{
    breaklinks = true,
    colorlinks = true,
    pdftitle = {}, 
    pdfauthor = {},
    pdfkeywords = {},
    linkcolor = blue,
    citecolor = blue,
    filecolor = black,
    urlcolor = magenta
}

\makeatletter 
\def\@author#1{\g@addto@macro\elsauthors{\normalsize%
    \def\baselinestretch{1}%
    \upshape\authorsep#1\unskip\textsuperscript{%
      \ifx\@fnmark\@empty\else\unskip\sep\@fnmark\let\sep=,\fi
      \ifx\@corref\@empty\else\unskip\sep\@corref\let\sep=,\fi
      }%
    \def\authorsep{\unskip,\space}%
    \global\let\@fnmark\@empty
    \global\let\@corref\@empty  
    \global\let\sep\@empty}%
    \@eadauthor={#1}
}
\makeatother

\begin{document}
\begin{sloppypar} 

\title{Magnetic properties of 3\textit{d}, 4\textit{d}, and 5\textit{d} transition-metal atomic monolayers in Fe/TM/Fe sandwiches: Systematic first-principles study}

\author{Justyn Snarski-Adamski \corref{cor1}}
\cortext[cor1]{Corresponding author} 
\ead{justyn.snarski-adamski@ifmpan.poznan.pl}

\author{Justyna Rychły}
\author{Mirosław Werwiński}

\address{Institute of Molecular Physics, Polish Academy of Sciences,\\  M. Smoluchowskiego 17, 60-179 Poznań, Poland}

\begin{abstract}
Previous studies have accurately determined the effect of transition metal point defects on the properties of bcc iron. The magnetic properties of transition metal monolayers on the iron surfaces have been studied equally intensively. In this work, we investigated the magnetic properties of the 3\textit{d}, 4\textit{d}, and 5\textit{d} transition-metal (TM) atomic monolayers in Fe/TM/Fe sandwiches using the full-potential local-orbital (FPLO) scheme of density functional theory. We prepared models of Fe/TM/Fe structures using the supercell method. We selected the total thickness of our system so that the Fe atomic layers furthest from the TM layer exhibit bulk iron-bcc properties. Along the direction perpendicular to the TM layer, we observe oscillations of spin and charge density. For Pt and W we obtained the largest values of perpendicular magnetocrystalline  anisotropy and for Lu and Ir the largest values of in-plane magnetocrystalline  anisotropy. All TM layers, except Co and Ni, reduce the total spin magnetic moment in the generated models, which is in good agreement with the Slater-Pauling curve. Density of states calculations showed that for Ag, Pd, Ir, and Au monolayers, a distinct van Hove singularity associated with TM/Fe interface can be observed at the Fermi level. 
\end{abstract}

\date{\today}

\maketitle

\section{Introduction}
\begin{figure}[t]
\centering
\includegraphics[clip, width=0.75\columnwidth]{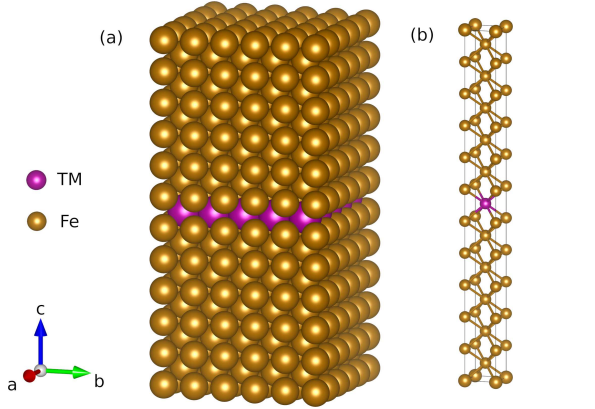}
\caption{\label{Warstwy}
A model of the crystal structure of the Fe/TM/Fe-sandwich system which crystallizes in a tetragonal structure [space group $P$4/$mmm$ (No. 123), $a$~=~$b$~=~2.83~\AA ~and $c$~=~31.13~\AA]. (a) Model showing the periodic multiplication of the unit cell, which is the subject of our consideration. (b) Unit cell of the Fe/TM/Fe-sandwich system. 
}
\end{figure}

Over the past decades, the magnetic thin films and layered structures have attracted considerable attention in theoretical and applied physics \cite{thersleff_towards_2017-1}. These systems exhibit novel physical phenomena such as enhanced magnetic moments, magnetocrystalline anisotropy (MAE), oscillatory interlayer coupling, and spin and charge-density waves \cite{fishman_spin-density_2001, turtur_magnetic_1994}. These quantum phenomena related to magnetism and spin-orbit coupling are also the subject of interest in spintronics. An important example of a spintronic device is the spin-transfer torque memory, which uses a spin-polarized tunneling current to switch magnetization \cite{slonczewski_current-driven_1996}. Although there is huge potential in creating magnetic multilayer structures, their control is difficult. Moreover, controlling and growing clean and sharp interfaces requires a large amount of labor and knowledge. Previous studies have predicted strong effects of 3$d$ TM monolayers on ferromagnetic moments on metallic overlayers \cite{fu_prediction_1985-1, li_giant_1991} or sandwich bilayers \cite{purcell_two-monolayer_1992}. For the systems considering the sandwiches model with Mn bilayers, oscillations of spin magnetic moments were observed \cite{purcell_two-monolayer_1992}.
The magnetic properties of 3$d$ TM on metallic substrates such as Pd(001) and Ag(001) show great similarity to the interactions of 3$d$ magnetic impurities in bulk alloys \cite{blugel_magnetic_1989}. For 4$d$ and 5$d$ monolayer on Ag(001) and Au(001) substrates, the magnetism has been explained as real effect of two-dimensional band structure \cite{blugel_two-dimensional_1992}. An exceptionally  large perpendicular MAE was found for Ir monolayer capped on Fe(001) surface \cite{odkhuu_extremely_2013-1}. There has also been interest in forcing ferromagnetic coupling in rare-earth-metal/Fe sandwiches systems using 3$d$ elements, which allow to obtaine huge magnetic moments of the order of 10~$\mu_{B}$/rare-earth atom \cite{sanyal_forcing_2010, autieri_recipe_2016}. Furthermore, the ferromagnetic coupling in the Fe/TM/Gd sandwich system with 4$d$ and 5$d$ metal spacers was found to be stronger than that with 3$d$ spacers \cite{autieri_systematic_2017}. In this work, we focused on the systems consisting of a Fe-bcc matrix and a transition metal monolayer, see Fig.~\ref{Warstwy}. 

\section{Computational details}
In this paper, the results of computations performed in the framework of density functional theory (DFT) are presented. The models of 3\textit{d}, 4\textit{d}, and 5\textit{d} transition-metal atomic monolayers in Fe/TM/Fe sandwiches were investigated using the full-potential local-orbital scheme (FPLO18.00-52) \citep{koepernik_full-potential_1999-1, eschrig_2_2004}. To model the systems, we used the supercell method \citep{edstrom_magnetocrystalline_2017}. The strucutral parameters of our models are $a$~=~$b$~=~2.83~\AA ~and $c$~=~31.13~\AA. The generalized gradient approximation (GGA) in the Perdew-Burke-Ernzerhof (PBE) form \cite{perdew_generalized_1996-1} was used. The lattice parameters were set as for bulk Fe-bcc and multiply 11 times in the $c$ direction, whereas the Wyckoff positions were optimized for each considered system using a spin-polarized scalar-relativistic approach, see Fig.~\ref{Warstwy}. The substitution of TM element was intended to create a single dopant monolayer in the center of this structure. Our calculation did not include optimization of lattice parameters.
A $12 \times 12 \times 1$ $k$-mesh was used for geometrical optimization and accuracy of forces was set as 10$^{-3}$ eV\,\AA$^{-1}$. An $80 \times 80 \times 10$ $k$-mesh was found to lead to well converged results of the magnetocrystalline anisotropy energies (MAE). The convergence criterion for the charge density was set as 10$^{-6}$. MAE was evaluated as a  difference between the fully-relativistic total energies calculated for quantization axes [100] and [001]. The applied full-potential approach plays an important role in the correct determination of the values of magnetic moments and MAE \cite{werwinski_magnetocrystalline_2018-1}. For Fe$_{22}$, the obtained result of the MAE value determines the accuracy of our calculations, as we would expect zero for such a system. Thus, the accuracy of our calculations is estimated at 1 \textmu eV/atom. 
The VESTA code was used to visualize the crystal structure \cite{momma_vesta_2008}.

\section{Results and discussion}
\begin{figure}[t]
\centering
\includegraphics[clip, width=0.9\columnwidth]{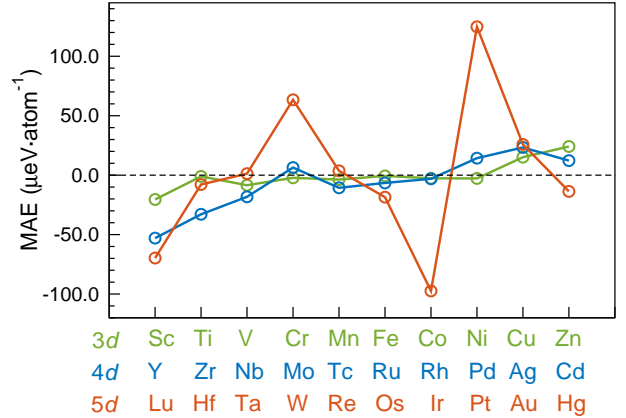}
\caption{\label{MAE}
Magnetocrystalline anisotropy energy (MAE) of Fe/TM/Fe-sandwiches for various 3$d$, 4$d$, and 5$d$ transition metal (TM) elements.
Calculations were performed with the FPLO18 using the PBE functional and supercell model. Considered unit cell include 21 atoms of Fe and 1 atom of TM. 
}
\end{figure}
\begin{figure*}[h!]
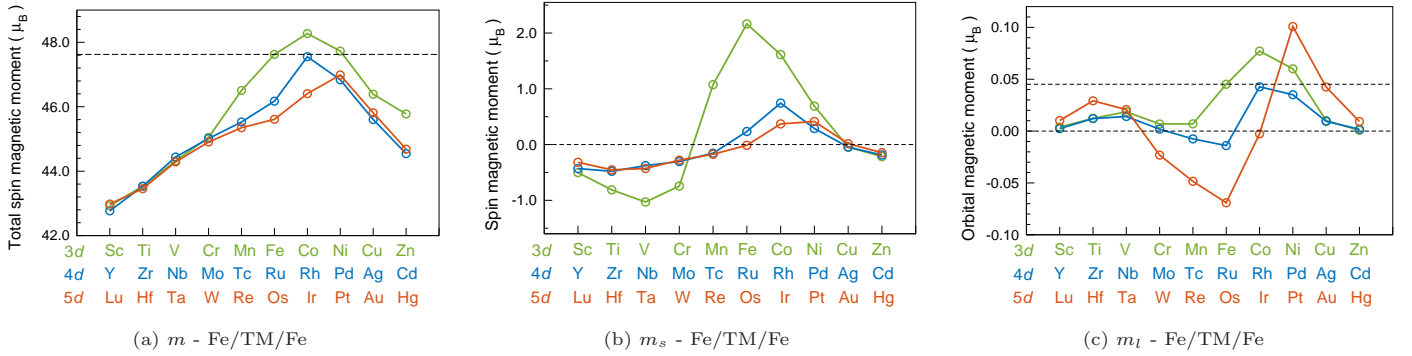

\centering
\subfloat[$m$ - Fe/TM/Fe]{\label{m}
\includegraphics[clip, width=0.31\textwidth]{Fe_TM_Fe_Total_Spin_Mom_3d_4d_5d.eps}}
\hfill
\subfloat[$m_s$ - Fe/TM/Fe]{\label{ms}
\includegraphics[clip, width=0.31\textwidth]{Fe_TM_Fe_ms_3d_4d_5d.eps}}
\hfill
\subfloat[$m_l$ - Fe/TM/Fe]{\label{ml}
\includegraphics[clip, width=0.31\textwidth]{Fe_TM_Fe_ml_3d_4d_5d.eps}}
\caption{
(\subref{m}) Total spin magnetic moment per formula unit, (\subref{ms}) spin magnetic moment as calculated for spin quantization axis along the distinguished axis [001] (easy axis), (\subref{ml}) orbital magnetic moment as calculated for spin quantization axis along the [001] (easy axis) for various 3$d$, 4$d$, and 5$d$ TM elements in the Fe/TM/Fe-sandwiches. Calculations were performed with FPLO18, using the PBE functional and supercell method.}
\end{figure*}
Since we considered all 3$d$, 4$d$, and 5$d$ transition metals in Fe/TM/Fe-sandwich models, we obtained a complete picture of the variations of MAE, total spin magnetic moment, and magnetic moments on TM, see Figs.~\ref{ms} and \ref{ml}. We can see that 5$d$ elements can lead to a more extreme MAE than the 3$d$ and 4$d$ elements due to the stronger spin-orbit coupling. Positive and negative MAE values indicate alignment of magnetic moments perpendicular and in-plane of the TM layer, respectively. For Pt and W we obtained the largest values of perpendicular magnetocrystalline  anisotropy and for Lu and Ir the largest values of in-plane magnetocrystalline  anisotropy. The largest value of MAE for Pt interlayer is not surprising considering the extremely high MAE values observed for the L1$_{0}$ FePt phase \cite{ristau_relationship_1999}. The previous calculations also showed an increase in the value of MAE for systems with W substitution \cite{edstrom_magnetic_2015-1, werwinski_magnetocrystalline_2018-1}.

Almost all TM elements, except Co and Ni, decrease the total spin magnetic moment relative to the Fe atom in the structures under our consideration, see Fig.~\ref{m}, which is in good agreement with the Slater-Pauling curve \cite{williams_generalized_1983}. We also see that almost all TMs, except Co, Ni, and Pt, contribute a lower orbital magnetic moment than Fe, see Fig.~\ref{ml}. Against this background, Pt stands out again, as it has the highest orbital moment among all the TMs considered. The orbital magnetic moment for bcc Fe calculated in this paper, which is 0.045~$\mu_{B}$, is underestimated relative to the experimental value of 0.085~$\mu_{B}$, which suggests that the calculated values of orbital magnetic moment are about twice underestimated. \cite{chen_experimental_1995-1}.
Calculations of the spin and orbital magnetic moment on the atomic monolayer of the transition metal show explicit trends with increasing atomic numbers. Analogous trends have been found for point substitutions in bulk-iron, both theoretically \cite{akai_nuclear_nodate, dederichs_ab-initio_1991-1} and experimentally \cite{wienke_determination_1991-1}.

\begin{center}
\begin{figure}[t]
\centering
\subfloat[]{\label{fig:SDW_CD}
\includegraphics[clip,width=0.95\columnwidth]{DW.eps}}
\hfill
\subfloat[]{\label{fig:fig1}
\includegraphics[clip,width=0.44\textwidth]{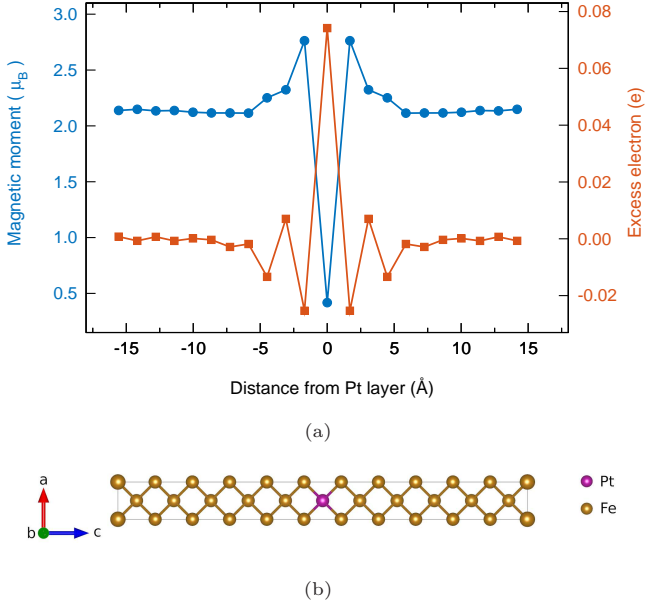}}
\caption{
(\subref{fig:SDW_CD}) Calculation of spin magnetic moment and charge oscillations in Fe/Pt/Fe-sandwich system. (\subref{fig:fig1}) A model of the crystal structure of Fe/TM/Fe-sandwich.}
\end{figure}
\end{center}

Therefore, we conducted a detailed case study for a system with Pt for which we obtained the maximum value of MAE.
In Fig.~\ref{fig:SDW_CD}, we have shown the oscillation of charge and spin magnetic moment in Fe/Pt/Fe-sandwich system. For Pt we observe  spin magnetic moment of 0.42 $\mu_{B}$ and for three Fe layers nearest to Pt the values are as follows: 2.76, 2.32, and 2.25 $\mu_{B}$. 
However, the Fe layers farthest from the Pt layer exhibit bulk Fe-bcc magnetic properties.

\begin{figure}[t]
\vspace{5mm}
\centering
\includegraphics[width = 0.95\columnwidth]{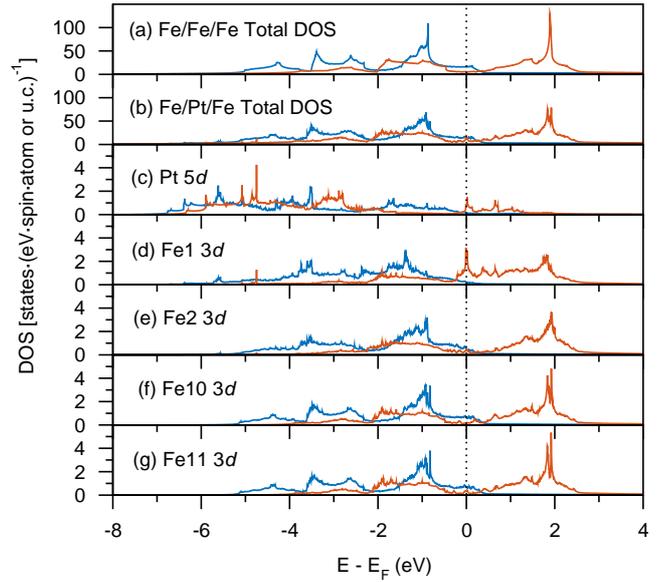}
\caption{\label{Fe_TM_Fe_DOS} 
Calculated densities of states (DOS) for the considered Fe/TM/Fe structure. (a) Fe$_{22}$ total DOS, (b) Fe/Pt/Fe total DOS, (c) Pt 5$d$ DOS, (d, e) Fe1 and Fe2 3$d$ DOS for the two Fe layers closest to the Pt layer, (f, g) Fe10 and Fe11 3$d$ DOS for the two Fe layers furthest from the Pt layer.
}
\end{figure}

We present the calculation of density of states to show what happens near the interface in the Fe/Pt/Fe-sandwich, see Fig.~\ref{Fe_TM_Fe_DOS}. In Fig.~\ref{Fe_TM_Fe_DOS}b we present the total DOS for the Fe/Pt/Fe system with the comparisons in Fig.~\ref{Fe_TM_Fe_DOS}a for Fe$_{22}$-bcc in the same model as here discussed (Fe/Fe/Fe-sandwich system). The comparison of Fe$_{22}$ DOS with Fe/Pt/Fe DOS shows great similarity and few differences resulting from about two additional electrons on the Pt 5$d$ shell compared to the Fe 3$d$ shell. The c-g panels show the contributions from the $d$ orbitals the two closest (Fe1, Fe2) and farthest (F10, Fe11) Fe layers in relation to the Pt layer. For the Pt 5$d$ band, one spin channel is partially empty and the other is completely filled. The adjacent Fe1 layer shows an increased spin polarization value ($m_s$ = 2.76 $\mu_{B}$). The DOS for Fe10 and Fe11, farthest from Pt layer, resemble the spectra observed for Fe$_{22}$. The van Hove singularity on the Fermi level observed for Fe/Pt/Fe-sandwich system comes from the Pt 5$d$ states. This narrow maximum at the Fermi level can have a significant impact on the electrochemical properties of the considered system. Unlike the band-types typical of valence band states, it can be interpreted as a characteristic of an atomic-like state of electrons. 

\section{Summary and conclusions}

We have presented the first-principles results for the Fe/TM/Fe-sandwich system. Almost all TM layers, except Co and Ni layers, reduce the total spin magnetic moment in the considered models, which agrees well with the Slater-Pauling curve. We observe that W and Pt monolayers induce strong perpendicular magnetocrystalline anisotropy, while Lu and Ir most strongly prefer in-plane anisotropy.
Our calculations indicate oscillation of spin magnetic moment and charge. We have also observed the van Hove singularity in the formed Fe/TM interface for TM~=~Ag, Pd, Pt, Ir, and Au.

\section*{Acknowledgements}
We acknowledge the financial support of the National Science Center Poland under the decision DEC-2019/35/O/ST5/02980 (PRELUDIUM-BIS 1) and DEC-2018/30/E/ST3/00267 (SONATA-BIS 8).
Part of the computations were performed on resources provided by the Poznan Supercomputing and Networking Center (PSNC).
We thank Paweł Leśniak and Daniel Depcik for compiling the scientific software and administration of the computing cluster at the Institute of Molecular Physics, Polish Academy of Sciences.

\section*{Appendix}
Table~\ref{tab:Fe/TM/Fe} provides the values shown in Figs.~\ref{MAE},~\ref{m},~\ref{ms}, and \ref{ml}.
\begin{center}
\begin{table*}
\caption{\label{tab:Fe/TM/Fe} Values of magnetocrystalline anisotropy energy [MAE (\textmu eV/atom)], total spin magnetic moment [$m$ ($\mu_B$ u.c.$^{-1}$)], spin magnetic moment [$m_s$ ($\mu_B$ u.c.$^{-1}$)] on transition metal (TM), and orbital magnetic moment [$m_l$~($\mu_B$~u.c.~$^{-1}$)] on TM for various 3$d$, 4$d$, and 5$d$ TM elements in the Fe/TM/Fe-sandwiches. }
\begin{tabular}{ccccc|ccccc|ccccc}
 \hline
 \hline
\multicolumn{5}{c}{3$d$}	&	\multicolumn{5}{c}{4$d$}	&	\multicolumn{5}{c}{5$d$}	\\																										
\hline																															
TM        &              MAE      &       $m$     &       $m_s$  &       $m_l$	&	TM        &              MAE      &       $m$     &       $m_s$  &       $m_l$	&	TM        &              MAE      &       $m$     &       $m_s$  &       $m_l$	\\																										
																															
\hline																															
Sc	&	-20.46	&	42.93	&	-0.50	&	0.0039	&	Y	&	-52.96	&	42.77	&	-0.43	&	0.0024	&	Lu	&	-69.73	&	42.98	&	-0.32	&	0.0102			\\
Ti	&	-1.06	&	43.53	&	-0.81	&	0.0123	&	Zr	&	-32.93	&	43.54	&	-0.48	&	0.0121	&	Hf	&	-7.88	&	43.46	&	-0.45	&	0.0292			\\
V	&	-8.47	&	44.32	&	-1.03	&	0.0186	&	Nb	&	-18.24	&	44.44	&	-0.38	&	0.0141	&	Ta	&	1.27	&	44.29	&	-0.43	&	0.0207			\\
Cr	&	-2.25	&	45.05	&	-0.74	&	0.0068	&	Mo	&	6.38	&	45.01	&	-0.30	&	0.0018	&	W	&	63.41	&	44.91	&	-0.28	&	-0.0232			\\
Mn	&	-3.69	&	46.50	&	1.07	&	0.0070	&	Tc	&	-10.62	&	45.53	&	-0.16	&	-0.0075	&	Re	&	3.65	&	45.35	&	-0.17	&	-0.0484			\\
Fe	&	-0.63	&	47.62	&	2.16	&	0.0451	&	Ru	&	-6.52	&	46.17	&	0.23	&	-0.0140	&	Os	&	-18.48	&	45.61	&	-0.01	&	-0.0692			\\
Co	&	-2.67	&	48.27	&	1.61	&	0.0771	&	Rh	&	-3.06	&	47.55	&	0.74	&	0.0427	&	Ir	&	-97.39	&	46.41	&	0.37	&	-0.0027			\\
Ni	&	-2.76	&	47.73	&	0.69	&	0.0601	&	Pd	&	14.28	&	46.83	&	0.29	&	0.0351	&	Pt	&	124.80	&	46.99	&	0.41	&	0.1008			\\
Cu	&	15.08	&	46.39	&	-0.04	&	0.0101	&	Ag	&	23.30	&	45.60	&	-0.05	&	0.0094	&	Au	&	25.82	&	45.81	&	0.02	&	0.0425			\\
Zn	&	24.08	&	45.77	&	-0.22	&	0.0008	&	Cd	&	12.22	&	44.54	&	-0.19	&	0.0017	&	Hg	&	-13.64	&	44.68	&	-0.15	&	0.0092			\\

\hline
\hline
\end{tabular}
\end{table*}
\end{center}

\bibliography{PM21}    

\end{sloppypar}
\end{document}